# Progressive amorphization of GeSbTe phase-change material under electron beam irradiation

Ting-Ting Jiang[a], Jiang-Jing Wang[a,b,*], Lu Lu[c], Chuan-Sheng Ma[c], Dan-Li Zhang[a], Feng Rao[d], Chun-Lin Jia[c,e], Wei Zhang[a,*]

[a]Center for Advancing Materials Performance from the Nanoscale, State Key Laboratory for Mechanical Behavior of Materials, Xi'an Jiaotong University, Xi'an 710049, China

[b]The School of Chemistry and Chemical Engineering, Yulin University, Yulin 719000, China

[c]The School of Electronic and Information Engineering, State Key Laboratory for Mechanical Behavior of Materials, Xi'an Jiaotong, University, Xi'an 710049, China

[d]College of Materials Science and Engineering, Shenzhen University, Shenzhen 518060, China

[e]Ernst Ruska-Centre for Microscopy and Spectroscopy with Electrons, Forschungszentrum Jülich GmbH, 52425 Jülich, Germany

*Corresponding author. Emails: wangjiangjing@stu.xjtu.edu.cn, wzhang0@mail.xjtu.edu.cn

**Abstract:**
Fast and reversible phase transitions in chalcogenide phase-change materials (PCMs), in particular, Ge-Sb-Te compounds, are not only of fundamental interests, but also make PCMs based random access memory (PRAM) a leading candidate for non-volatile memory and neuromorphic computing devices. To RESET the memory cell, crystalline Ge-Sb-Te has to undergo phase transitions firstly to a liquid state and then to an amorphous state, corresponding to an abrupt change in electrical resistance. In this work, we demonstrate a progressive amorphization process in $GeSb_2Te_4$ thin films under electron beam irradiation on transmission electron microscope (TEM). Melting is shown to be completely absent by the *in situ* TEM experiments. The progressive amorphization process resembles closely the cumulative crystallization process that accompanies a continuous change in electrical resistance. Our work suggests that if displacement forces can be implemented properly, it should be possible to emulate symmetric neuronal dynamics by using PCMs.

Artificial intelligence, big data analytics and other data-intensive applications are changing our lives rapidly and profoundly. They, however, also pose a significant challenge to data storage and processing to the current computing architecture, which physically separates storage units from processing units. The extensive data shuffling between these units over band-width limited interconnects leads to a fundamental barrier in improving the computing efficiency. The emerging neuromorphic computing devices[1-7] hold the promise to break this barrier by unifying storage with processing in a single cell. Phase change materials (PCMs) based random access memory (PRAM) is one of the leading candidates for this application[6-14].

PRAM is technologically mature and has entered the global memory market as Storage-Class Memory (SCM) recently[15], filling the performance gap between dynamic random access memory (DRAM) and flash memory based solid state hard drive (SSD). The basic principle of PRAM is to exploit the large contrast in electrical resistance, and the rapid and reversible transitions between two solid states of PCMs, i.e. a disordered amorphous state and an ordered crystalline state[16]. Among the explored PCMs, Ge-Sb-Te compounds along the GeTe-$Sb_2Te_3$ pseudobinary line[17-25], such as $Ge_2Sb_2Te_5$ and $GeSb_2Te_4$, are the most widely studied materials. Upon crystallization, amorphous (amor-) Ge-Sb-Te compounds form a metastable cubic rocksalt (cub-) phase[26-35].

One key attribute of PCM for neuromorphic computing is that its crystallization process can be accomplished by multiple narrow voltage pulses in a cumulative mode, corresponding to a partial SET process of PRAM[7]. The change in electrical conductance of the memory cell is then dependent on the previous excitation. This history-dependent behavior of electrical resistance classifies PRAM as memristors[36,37]. To RESET the crystallized PRAM cell, a high voltage pulse is applied to melt down the crystalline state, and the amorphous state is obtained upon rapid cooling. This melt-quench procedure leads to an abrupt RESET process, hindering the emulation of symmetric neuronal dynamics using PCMs[7].

Here, we demonstrate a progressive amorphization process in $GeSb_2Te_4$ (short as GST in the following) thin films under electron beam (E-beam) irradiation on transmission electron microscope (TEM). TEM is an important technique to assess the microstructure and elemental distribution of materials, while as high energy particle beams, electron beams can also be used to cause temporary or permanent changes of the specimen structure, e.g. to induce phase transitions of PCMs. In this work, *in situ* E-beam irradiation experiments were performed on a JEOL JEM-2100F TEM under 200 keV and a FEI Titan G2 aberration-corrected TEM under 300 keV.

GST films of ~80 nm-thick were deposited with magnetron sputtering on ultra-thin carbon film (~5 nm) TEM grids, and were covered by an electron-transparent $ZnS-SiO_2$ layer to prevent oxidization.

The GST thin films were in amorphous phase upon sputtering. One set of samples was irradiated using electron beams, while another set of samples was annealed at 150 °C for 1 hour in argon atmosphere (flow rates of 1200 sccm) in a regular tube furnace for comparison. The thermal-based and E-beam-based phase transitions of GST are depicted in Fig. 1. Upon heating, amor-phase crystallizes into cub-phase at ~150 °C, while to induce the reversed transition, the cub-phase needs to be melted at ~650 °C and then rapidly cooled down to room temperature. In contrast, direct and reversible phase transitions between amor- and cub-GST can be obtained by manipulating the beam intensity, accelerating voltage and irradiation time of TEM.

We start our irradiation experiment on the JEOL JEM-2100F TEM with a fixed accelerating voltage of 200 keV and a set beam intensity of $6.0 \times 10^{23}$ e m$^{-2}$ s$^{-1}$ (dose, $5.4 \times 10^{8}$ e nm$^{-2}$). The amorphous nature of the initial phase is confirmed by the halo rings of the selected area electron diffraction (SAED) patterns (Fig. 2a). Multiple crystal nuclei appear after 5 min irradiation, leading to visible changes in contrast of the bright-field image and diffraction rings in the SAED pattern (Fig. 2b). Upon further irradiation to 15 min, randomly oriented crystal grains with average size of 10-20 nm are observed, and the corresponding SAED pattern shows bright and sharp diffraction rings with cub-phase characteristics (Fig. 2c), which are comparable with those of the thermally-annealed sample (Fig. 2f). This E-beam-induced crystallization has been observed consistently in PCMs[38-41]. By reducing the beam intensity, crystallization was effectively prohibited. For example, no obvious change in amorphous GST could be observed in JEOL JEM-2100F TEM with beam intensity lower than $1.6 \times 10^{23}$ e m$^{-2}$ s$^{-1}$ within 90 min[41] (this beam intensity value at irradiation area corresponds to the value of $1.0 \times 10^{12}$ e m$^{-2}$ s$^{-1}$ at screen shown in Ref. 41). Such time window is sufficiently long for TEM measurements to assess the structural details amorphous GST[42].

It was concluded that in addition to heating effects, the displacement damage by knock-on collisions of E-beams[43] also plays a role in triggering crystallization in GST thin films[38]. Significant atomic displacement was also observed in crystalline GST by focusing the electron beam to a small area for extensive irradiation[44,45]. Following the same strategy, we made the electron beams more focused to increase the probability of knock-on collisions, and to check if this kinetic effect could be strong enough to induce amorphization like the case of ion beam irradiations[46,47]. Indeed, parts of the irradiation-crystallized thin film transformed into amorphous phase (Fig. 2d) after 45 min E-beam irradiation, and after 75 min, almost all the cubic grains vanished and dim halo rings appeared in the SAED pattern (Fig. 2e). The measured beam intensity was $1.1 \times 10^{24}$ e m$^{-2}$ s$^{-1}$ (dose, $4.95 \times 10^{9}$ e nm$^{-2}$). The same irradiation experiment was also carried out on the thermally-annealed cubic-phase sample, and similar amorphization process was observed over ~80 min. No sign of amorphization can be observed, if the electron beams are less focused (beam intensity below ~8.0

× $10^{23}$ e m$^{-2}$ s$^{-1}$). The irradiation-induced amorphous GST can be further crystallized upon electron beam irradiation.

It is noted that the observed irradiation-induced amorphization process is continuous, and no abrupt change in structural patterns could be observed, which indicates a non-thermal phase transition to the amorphous state, bypassing the melt-quench process. To gain additional support of a non-thermal-dominated transition, we made a rough estimate of the temperature increase in the thin film of GST following the equation[48], $\Delta T = \frac{I}{\pi k e}\left(\frac{\Delta E}{d}\right)\ln\frac{b}{r_0}$, in which *e, I, $r_0$, k, and b* are the electron's charge, beam current, beam radius, thermal conductivity and sample radius, respectively. $\Delta E$ is the total energy loss per electron in a sample of thickness *d*. The estimated maximum temperature rise is ~220 °C, which is well below the melting point of GST (~650 °C). In fact, GST thin films will evaporate quickly above 450 °C, and it is not feasible to observe the melting process at ~650 °C in unencapsulated GST thin films.

To explore the structural details during the E-beam induced amorphization process, high resolution TEM (HRTEM) characterizations were carried out. Fig. 3a shows the initial stage. The center target grain shows [011]-orientation of cubic lattice, as evidenced by its fast Fourier transform (FFT) pattern. The FFT pattern corresponds to the white dashed box in the HRTEM image. The inset depicts a zoom-in view of the area inside the red box, showing that the inter-plane spacing of (111) is 3.6 Å and the angle between ($1\bar{1}1$) and ($\bar{1}\bar{1}1$) is 71°. Upon E-beam irradiation over 40 min (Fig. 3b), the original [011]-oriented grain (marked by dashed yellow line) split into four cub-phase grains with different orientations. The orientation of the primary grain (solid lines in Fig. 3b) remained unchanged. At this stage, amorphization also took place, which reduced the size of the primary grain, as indicated by arrows in Fig. 3b. Another 20 min irradiation amorphized the major parts of the irradiation area, and the initial grain further split into several smaller parts (Fig. 3c). Finally, all the cub-phase grains disappeared and the whole irradiation area turned amorphous after 80 min irradiation, as confirmed by the corresponding FFT pattern (Figure 3d).

This direct cub- to amor-GST transition path under E-beam irradiation is in a stark contrast with the conventional transition path induced by thermal melt-quenching. Clearly, no abrupt structural change to liquid state could be observed during the *in situ* TEM experiments. Instead, the E-beam induced amorphization undergoes gradual structural rearrangement locally, which can be attributed to the knock-on effects of E-beams. In general, for electron illumination of thin specimen, a higher accelerating voltage results in less effects of specimen heating, but stronger kinetic effects of knock-on collisions[49,50]. For further confirmation, we performed *in situ* experiments on a FEI Titan G2 TEM operated at 300 keV. It was found that the beam intensity in this TEM can reach ~1.9 × $10^{32}$ e m$^{-2}$ s$^{-1}$ (dose, ~2.28 × 10$^{17}$ e nm$^{-2}$). Starting from cub-GST, amorphization already proceeded after 5 min irradiation, as shown in Fig. 4a. As the irradiation time increased, amorphous region expanded gradually and rapidly (Fig. 4b and 4c). The faster amorphization on the FEI Titan G2 TEM than on the JEOL JEM-2100F TEM can be understood due to the irradiation by the beam with much higher intensity and at the higher accelerating voltage. As stated above, the knock-on collision effects should be the major source for amorphization, as heating effects should easily evaporate the unencapsulated thin film.

In summary, we have demonstrated an effective approach to achieve reversible and direct crystallization and amorphization of GeSb$_2$Te$_4$ in a progressive manner by means of E-beam irradiation. The chief TEM parameters for these transitions are found to be accelerating voltage, beam intensity, and irradiation time. The *in situ* irradiation experiments provide a real-time and real-space view of progressive structural evolution between the two solid state phases, where melting is completely absent. The knock-on collision effect of E-beams drives this non-thermal amorphization process. We note that the displacement forces induced by electron or ion beam irradiations are not likely to be implemented in electronic devices, while mechanical forces and strains induced by e.g. piezoelectric materials[51] appear to be a more suitable approach[52]. A previous work on silicon nanopillar also shows a continuous and progressive non-thermal amorphization process under uniaxial strain[53]. Similar experiments are anticipated to be carried out in PCMs. If displacement forces can be properly implemented, symmetric neuronal dynamics could possibly be emulated using PCMs.

**Acknowledgements**
W.Z. thanks the support of National Natural Science Foundation of China (61774123), 111 Project 2.0 (BP2018008), the Science and Technology Department of Jiangsu Province (BK20170414), and the Young Talent Support Plan and of Xi'an Jiaotong University. F.R. gratefully thanks the National Natural Science Foundation of China (61622408), the Major Provincial Basic Research Program of Guangdong (2017KZDXM070), and the Science and Technology Foundation of Shenzhen (JCYJ20180507182248605). The authors also acknowledge the support by the International Joint Laboratory for Micro/Nano Manufacturing and Measurement Technologies of Xi'an Jiaotong University.

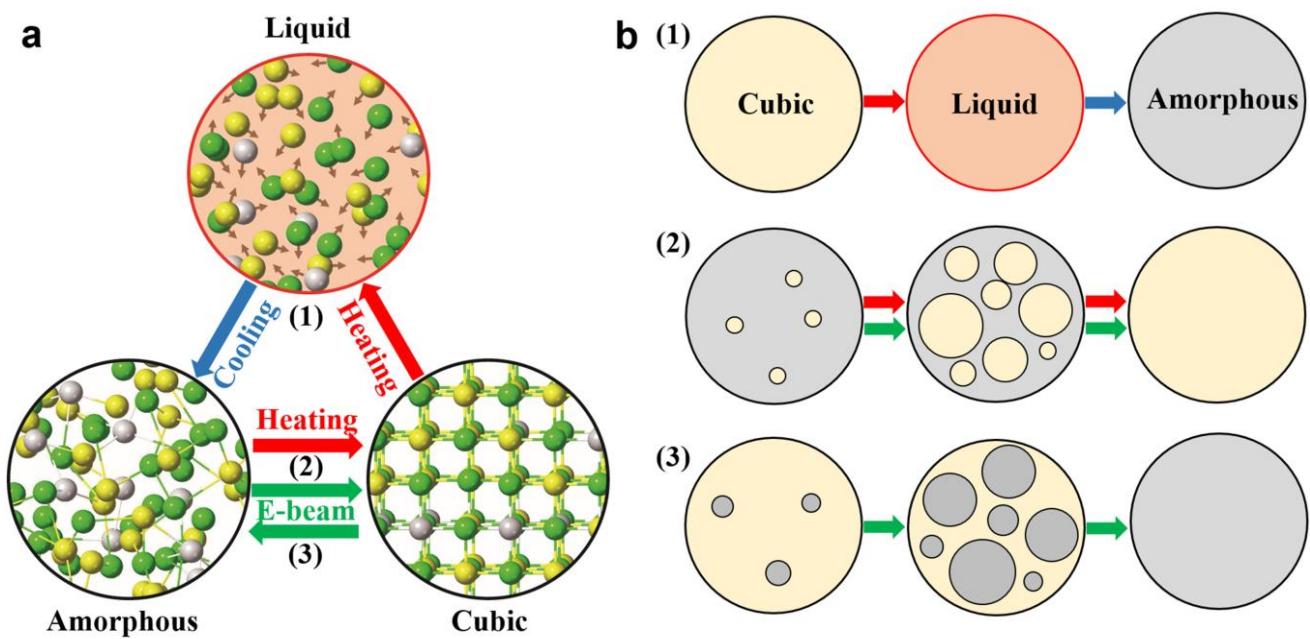

**Figure 1.** Phase transition paths of GST. **a** Crystallization can be accomplished by heating, while to trigger amorphization by thermal power, melting and subsequent rapid cooling are necessary. As regards electron beam (E-beam) irradiation, direct and reversible solid-state transitions can be achieved. **b** The sketches of (1) abrupt amorphization by melt-quenching, (2) progressive crystallization upon heating or by E-beam irradiation, and (3) progressive amorphization by E-beam irradiation.

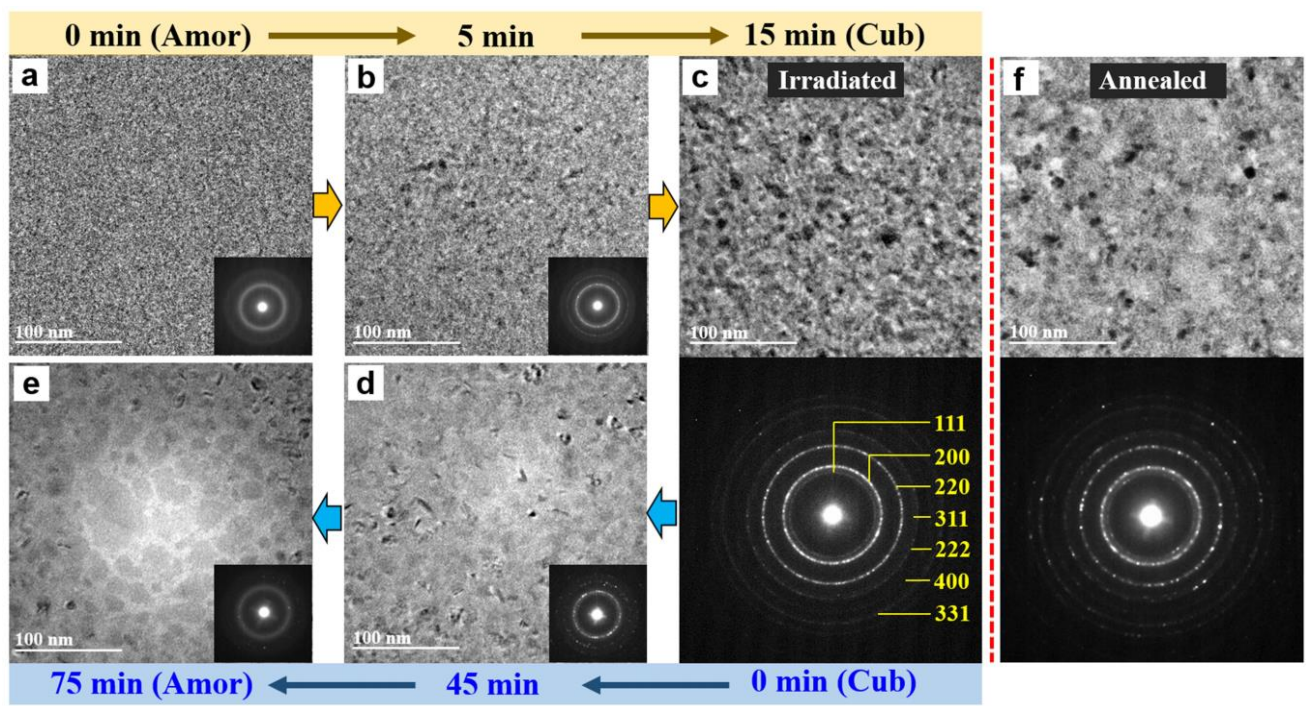

**Figure 2.** The bright-field TEM images and the corresponding SAED patterns of GST, recorded under electron beam irradiation. **a-c** Crystallization. **c-e** Amorphization. The critical parameter in inducing the opposite transitions is beam intensity, i.e. $6.0 \times 10^{23}$ e m$^{-2}$ s$^{-1}$ (dose, $5.4 \times 10^8$ e nm$^{-2}$) for crystallization, and $1.1 \times 10^{24}$ e m$^{-2}$ s$^{-1}$ (dose, $4.95 \times 10^9$ e nm$^{-2}$) for amorphization. The accelerating voltage is the same, 200 keV. **f** TEM image and the corresponding SAED pattern of GST thin films annealed at 150 °C for 1 hour, showing the comparable features of the E-beam induced cubic GST sample shown in **c**.

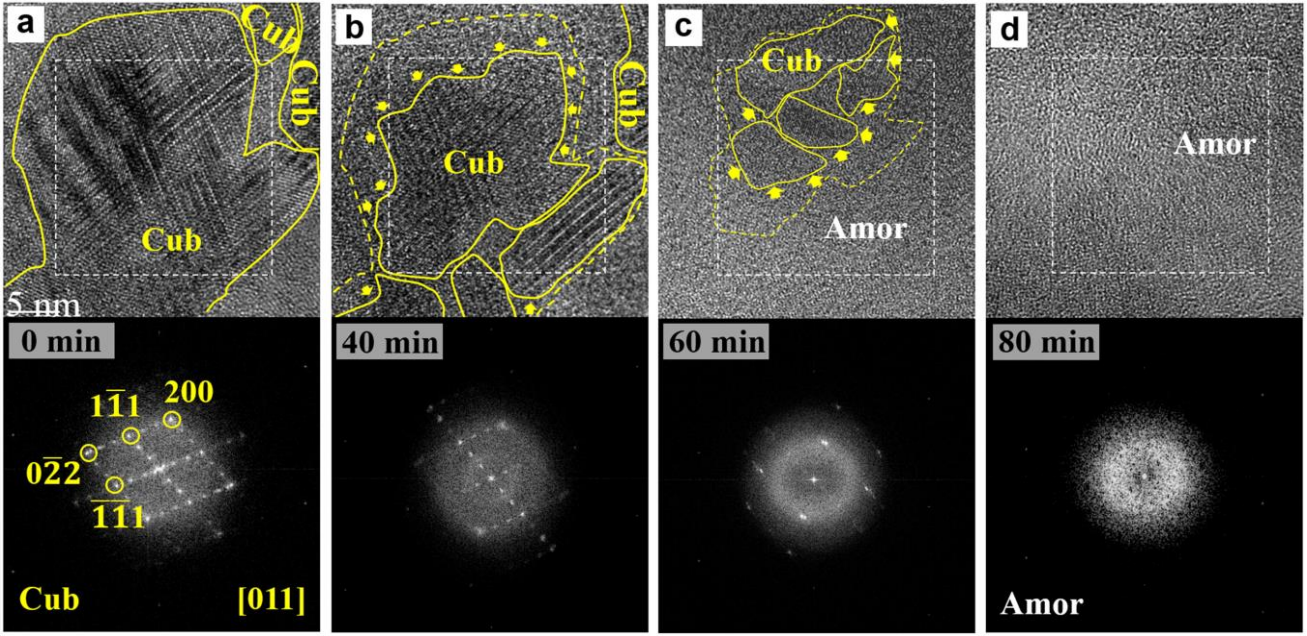

**Figure 3**. *In situ* amorphization of cubic GST under 200 keV E-beam irradiation. **a-d** The snapshots of phase transition from cub- to amor-phase as a function of irradiation time, showing the stages for separation and shrinking of the cub-phase grains. After 80 min irradiation, all the cub-phase grains turn amorphous completely. The image areas in the dashed white boxes in the bright-field images are used for the calculations of fast Fourier transform (FFT) patterns.

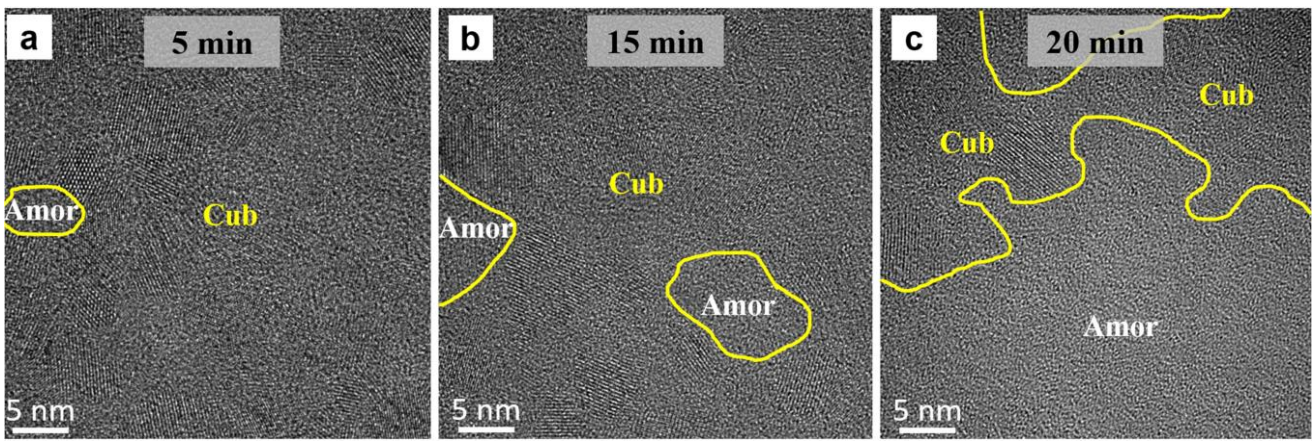

**Figure 4.** *In situ* amorphization of cub-GeSb$_2$Te$_4$ under 300 keV e-beam irradiation with a beam intensity ~1.9 × 10$^{32}$ e m$^{-2}$ s$^{-1}$ (dose, ~2.28 × 10$^{17}$ e nm$^{-2}$). **a-c** The snapshots of phase transition from cub- to amor-phase as a function of irradiation time. Higher accelerating voltage and higher beam intensity lead to a more rapid amorphization process.